\documentclass[onecolumn,floatfix,aps,showpacs,nofootinbib]{revtex4}
\usepackage{amssymb,amsmath}
\usepackage{amscd}
\usepackage{subfigure}
\usepackage[dvips]{graphics,color}
\usepackage{epsfig}\usepackage{float}
\newcommand{\hko}{\hookrightarrow}

\newcommand{\bec}{\begin{center}}
\newcommand{\eec}{\end{center}}

\newcommand{\bea}{\begin{array}}
\newcommand{\ear}{\end{array}}

\newcommand{\bfr}{\begin{flushright}}

\newcommand{\efr}{\end{flushright}}
\newcommand{\noi}{\noindent}\newcommand{\Ra}{\rightarrow}
\newcommand{\me}{\frac{1}{2}}

\newcommand{\cl}{{\mt{C}}\ell}
\newcommand{\RR}{\mathbb{R}}\newcommand{\op}{\oplus}
\newcommand{\HH}{\mathbb{H}}\newcommand{\PP}{\mathbb{P}}

\newcommand{\ot}{\otimes}
\newcommand{\la}{\Lambda}
\newcommand{\bege}{\begin{equation}}
\newcommand{\enge}{\end{equation}}
\newcommand{\w}{\wedge}
\newcommand{\g}{\gamma}

\newcommand{\ri}{\rightarrow}

\newcommand{\ty}{\RR\oplus\RR^3}
\newcommand{\si}{\sigma}

\newcommand{\beq}{\begin{eqnarray}}\newcommand{\benu}{\begin{enumerate}}\newcommand{\enu}{\end{enumerate}}
\newcommand{\eeq}{\end{eqnarray}}
\newcommand{\mt}{\mathcal}

\newcommand{\vv}{{\bf v}}
\newcommand{\ee}{{\bf e}}
\newcommand{\uu}{{\bf u}}

\newcommand{\ww}{{\bf w}}
\newcommand{\CC}{\mathbb{C}}

\newcommand{\KK}{\mathbb{K}}

\newcommand{\ZZ}{\mathbb{Z}}

\newcommand{\vbn}{\blacktriangleleft}
\newcommand{\vvn}{\blacktriangleright}

\newcommand{\clt}{{\mt{C}}\ell_{3,0}}
\newcommand{\cle}{{\mt{C}}\ell_{1,3}}

\newcommand{\BA}{\breve{A}}

\newcommand{\BB}{\breve{B}}

\newcommand{\bx}{\begin{pmatrix}}
\newcommand{\ex}{\end{pmatrix}}
\newcommand{\vcx}{\varepsilon}

\newcommand{\mmb}{{\mathfrak{b}}}
\newcommand{\mmg}{{\mathfrak{g}}}

\begin{document}

\title{$\kappa$-deformed Poincar\' e algebras and quantum Clifford-Hopf algebras}
\author{Rold\~ao da Rocha}
\email{roldao.rocha@ufabc.edu.br}
\affiliation{Centro de Matem\'atica, Computa\c c\~ao e Cogni\c
c\~ao,
Universidade Federal do ABC, 09210-170, Santo Andr\'e, SP, Brazil}
\author{Alex E. Bernardini}
\email{alexeb@ufscar.br}
\affiliation{Departamento de F\'{\i}sica, Universidade Federal de S\~ao
Carlos PO Box 676, 13565-905, S\~ao Carlos, SP, Brazil}
\author{Jayme Vaz, Jr.}
\email{vaz@ime.unicamp.br}
\affiliation{Departamento de Matem\'atica Aplicada, IMECC Unicamp, CP 6065, 13083-859, Campinas, SP, Brazil.}
\begin{abstract}
The Minkowski spacetime quantum Clifford algebra structure associated with the conformal group and the Clifford-Hopf
alternative $\kappa$-deformed quantum Poincar\' e algebra is investigated in the Atiyah-Bott-Shapiro mod 8 theorem context. The resulting algebra is equivalent
to the deformed anti-de Sitter algebra $\mathcal{U}_q$($\mathfrak{so}$(3,2)),
when the associated Clifford-Hopf algebra is taken into account, together with
the associated quantum Clifford algebra and a (not braided) deformation of the periodicity
Atiyah-Bott-Shapiro theorem.
 \end{abstract}
\maketitle
\section{Introduction}
Conformal symmetry represents the fundamental spacetime symmetry,
and it contains the Poincar\'e and de Sitter geometries as particular cases,
besides describing massless particles and field symmetries.
In order to investigate modifications of the relativistic kinematics at sufficiently high energy,
quantum deformations of the Poincar\'e algebra \cite{lu2,majid,lu3,lu1,6gi,7gi,8gi,9gi} were introduced and followed by the  doubly special relativity (DSR), which contains two observer-independent parameters --- the light velocity
and the Planck length \cite{ame1,bruno}.
The DSR framework coincides with the algebraic structure of the Poincar\'e algebra $\kappa$-deformation, where
the deformation parameter $\kappa$ is related to the Planck mass.
The DSR formalism can indeed be introduced using the quantum $\kappa$-Poincar\'e algebra
\cite{lu2,kappa,majid}, which presents a deformation parameter $\kappa$
of dimension of mass \cite{glik,bruno}.
One of the basic physical predictions of DSR, with the $\kappa$-Poincar\'e
algebra as the symmetry algebra of the theory, is
the existence of an observer-independent fundamental mass scale \cite{bruno,now1}.
In addition, in $D=4$ Poincar\'e algebra there is a type of quantum
deformation with the mass-like parameter $\kappa$ \cite{g1,g2,g3,g4,g5,h1}.
The introduction of the deformation parameter $\kappa$ leads to the arising of the fundamental mass on fundamental geometrical level, and the corresponding deformations of conformal algebras introduce an original case of quantum deformations of Lie
algebras, generalizing the nonstandard deformation of $\mathfrak{sl}$(2) \cite{q1,q2,q3,q4,q5}.
Moreover, deformations of relativistic symmetries in the framework of quantum groups have also been considered \cite{drin, fadeev, majid1}, and $\kappa$-deformations have brought wide applications in physics \cite{g1,g3,g4,e3,e4,h1,h2,h3,h31,h32,i1,i3}.
It has introduced a mass-like deformation parameter $\kappa$ related to the Planck mass as well as to quantum gravity corrections.
In particular, in the framework of Lorentz-invariance violation, modifications to the fermionic particle equation of motion were introduced, and aspects on deformed dilation transformations and some prominent applications were investigated \cite{alex1,alex2,ma3}.

A recent classification of deformed Poincar\'e groups has been presented \cite{zumino} based on the Lorentz group deformations.
In connection with deformed Minkowski spaces \cite{61,i2,7gi,q3}, the quantum deformed $\kappa$-Poincar\'e algebra \cite{kos} can thus be obtained through a nonstandard contraction of the deformed anti-de Sitter algebra $\mathcal{U}_q$($\mathfrak{so}$(3,2)).

In this paper we formulate the $\kappa$-Poincar\'e algebra as a quantum Clifford-Hopf algebra,
using the Wick isomorphism that relates quantum Clifford algebras to their respective standard Clifford algebras.
The main aspects of quantum Clifford algebras are reviewed in Section III, where we point out
some developments by Hestenes, Oziewicz, Lounesto, Ab\l amowicz, and Fauser \cite{39, 9611, 9908, 5aa, 6aa}.
Quantum Clifford algebras have been widely investigated, relating the $\ZZ_n$-graded Clifford algebra structure to $q$-quantization.
Some physical systems \cite{9611} are regarded, and to explore the formal point of view, see, e.g., \cite{6aa}.
Also, the $q$-symmetry and Hecke algebras can
be described within the quantum Clifford algebra context \cite{9908}. Fauser asserts that this
structure should play a major role in the discussion of the Yang-Baxter equation,
the knot theory, the link invariants and in other related fields which are crucial
for the physics of integrable systems in statistical physics \cite{5aa, 6aa}, where in some cases bivectors satisfy minimal polynomial equations of the Hecke type \cite{6aa}.
In addition, there are other germane applications concerning this formalism, for instance the structure theory of Clifford algebras over arbitrary
rings \cite{a36}, and the arithmetic theory of Arf invariants and the Brauer-Wall groups.
It was shown that due to central extensions the ungraded
bivector Lie algebras turn into Kac-Moody and Virasoro algebras and, as
it is also shown in \cite{9908}, to some $q$-deformed algebras.
Automorphisms generated by non-isotropic vectors can give rise to infinite dimensional Coxeter
groups \cite{cox}, affine Weyl groups, connected to $\ZZ_n$-graded quantum
Clifford algebras.

Motivated by these considerations, the main aim of this paper is to evince the Clifford-Hopf character associated with
quantum $\kappa$-deformed Poincar\'e algebras, as a consequence of a specific deformation of the conformal Clifford algebra
$\cl_{2,4}$ into its associated quantum Clifford algebra. Once the algebra of conformal transformations is derived, comprised solely in terms of
the real vectors in Minkowski spacetime, the related periodicity theorem of quantum Clifford algebras is considered, regarding the deformed tensor product that is not braided by construction, in full compliance with the Hopf underlying algebraic structure of quantum Clifford algebras.

This paper is organized as follows:  after presenting some algebraic preliminaries in Section II, in Section III
we briefly review some introductory aspects on quantum Clifford-Hopf algebras, in particular their $\ZZ_2$-graded co-commutative structure, and the algebraic sector accomplished by the quantum Clifford algebras. Also the Wick isomorphism is introduced. In Section IV the spacetime algebra $\cle$ and
  the algebras $\cl_{4,1}$ and $\cl_{2,4}$ are briefly introduced, in order to show
   the well known isomorphism between the associated Spin group and the fourfold covering of the special conformal transformations group.
   All the conformal maps are recalled in terms of the spacetime algebra and the Atiyah-Bott-Shapiro periodicity theorem. The algebraic aspects of the conformal transformations are deeply investigated, where the Lie algebra of the associated groups is reviewed, together with the
fact that their elements are 2-forms, and the Poincar\'e algebra is obtained in this context, also reviewing some general aspects.
Finally in Section V the $\kappa$-Poincar\'e algebra is described as a quantum Clifford-Hopf algebra, together with its Lie algebra character, where the conformal group and an alternative $\kappa$-deformed Poincar\'e algebra are evinced solely in terms
of elements of the spacetime algebra  $\cle$. The conformal transformations and $\kappa$-deformed Poincar\'e algebras, and a quantum $\kappa$-deformed Poincar\'e symmetry are formulated  together with the respective Hopf algebra relations, in the context of quantum Clifford algebras and
accomplished by their associated Wick isomorphism.

\section{Preliminaries}
\label{1}
Let $V$ be a finite $n$-dimensional real vector space and $V^*$ denotes its dual.
We consider the tensor algebra $\oplus_{i=0}^\infty T^i(V)$ from which we
restrict our attention to the space $\Lambda(V) = \oplus_{k=0}^n\Lambda^k(V)$ of multivectors over $V$. $\Lambda^k(V)$
denotes the space of antisymmetric
 $k$-tensors, isomorphic to the  $k$-forms vector space.  Given $\psi\in\Lambda(V)$, $\tilde\psi$ denotes the \emph{reversion},
 an algebra antiautomorphism
 given by $\tilde{\psi} = (-1)^{[k/2]}\psi$ ([$k$] denotes the integer part of $k$). $\hat\psi$ denotes
the \emph{main automorphism or graded involution},  given by
$\hat{\psi} = (-1)^k \psi$. The \emph{conjugation} is defined as the reversion followed by the main automorphism.
  If $V$ is endowed with a non-degenerate, symmetric, bilinear map $g: V^*\times V^* \rightarrow \RR$, it is
possible to extend $g$ to $\la(V)$. Given $\psi=\uu^1\w\cdots\w \uu^k$ and $\phi=\vv^1\w\cdots\w \vv^l$, for $\uu^i, \vv^j\in V^*$, one defines $g(\psi,\phi)
 = \det(g(\uu^i,\vv^j))$ if $k=l$ and $g(\psi,\phi)=0$ if $k\neq l$. The projection of a multivector $\psi= \psi_0 + \psi_1 + \cdots + \psi_n$,
 $\psi_k \in \la^k(V)$, on its $p$-vector part is given by $\langle\psi\rangle_p$ = $\psi_p$.
 Given $\psi,\phi,\xi\in\Lambda(V)$, the  {\it left contraction} is defined implicitly by
$g(\psi\lrcorner\phi,\xi)=g(\phi,\tilde\psi\w\xi)$.
 For $a \in \RR$, it follows that
 ${\bf v} \lrcorner a = 0$. Given $\vv\in V$, the Leibniz rule
${\bf v}\lrcorner (\psi \w \phi) = ({\bf v} \lrcorner \psi) \w \phi + \hat\psi \w ({\bf v} \lrcorner \phi)$ holds. The
 {\it right contraction} is analogously defined
$g(\psi\llcorner\phi,\xi)=g(\phi,\psi\w\tilde\xi)$
 and its associated Leibniz rule $
(\psi \w \phi) \llcorner {\bf v} = \psi \w (\phi \llcorner {\bf v}) + (\psi \llcorner {\bf v}) \w \hat\phi
$ holds. Both contractions are related by
${\bf v} \lrcorner \psi = -\hat\psi \llcorner {\bf v}$.
The Clifford product between $\ww\in V$ and $\psi\in\la(V)$ is given by $\ww\psi = \ww\w \psi + \ww\lrcorner \psi$.
 The Grassmann algebra $(\la(V),g)$
endowed with the Clifford  product is denoted by $\cl(V,g)$ or $\cl_{p,q}$, the Clifford algebra associated with $V\simeq \RR^{p,q},\; p + q = n$.
In what follows $\RR,\CC$ and $\HH$ denote respectively the real, complex and quaternionic  fields. The vector space $\Lambda_k(V)$ denotes
the space of the $k$-vectors.

\section{Clifford-Hopf Algebras and Quantum Clifford Algebras}

The Clifford-Hopf algebra
can be introduced when a compatible co-algebra structure and an antipode endow the standard Clifford algebra structure.
Denoting the unit map embedding the real or complex field $\KK$ into the algebra
with $\eta: \KK\Ra \cl(V,g)$, the co-algebra structure is then given by a co-product $\Delta: \cl(V,g)\Ra\cl(V,g)\ot\cl(V,g)$ and a co-unit $\epsilon: \cl(V,g)\Ra\KK$, which arise naturally when there is a functorial dualization of  the algebra structure \cite{38,majid1,5aa,6aa}. The compatibility of the algebra and the co-algebra structures requires the co-product and the co-unit to be algebra homomorphisms:
$\epsilon({\rm Id}) = 1$, $\epsilon(\psi \phi) = \epsilon(\psi) \epsilon(\phi)$ --- the juxtaposition denoting the Clifford product ---
and also the co-product must satisfy
\beq
\Delta({\rm Id}) = {\rm Id} \otimes {\rm Id}, \qquad \Delta(\psi) = \psi\otimes {\rm Id} + {\rm Id}\otimes \psi,\qquad \Delta(\psi\phi) = \Delta(\psi)\Delta(\phi).\eeq Define also the antipode $S: \Lambda^p(V)  \Ra \Lambda^p(V)$ as
$S(\psi_p) = (-1)^p \psi_p$, where $\psi_p\in\Lambda^p(V)$. The antipode can be also extended by linearity to $\Lambda(V)$ and
can be identified to the graded involution of the underlying Grassmann
algebra. Denoting the Clifford application by $\gamma: V \Ra \cl(V,g)$ --- satisfying $\gamma(u)\gamma(v) + \gamma(v)\gamma(u) = 2g(u,v),\, \forall u,v\in V$ --- the 6-tuple  $(\Lambda(V), g, \gamma, \epsilon, \Delta, S)$
is the Clifford-Hopf algebra, which is unique up to the isomorphism given by the universality property \cite{38}.
From the definitions above all axioms for a
Hopf algebra are satisfied, together with the antipode axioms $
\wedge \circ (S \ot {\rm Id}) = \eta\circ \epsilon$ and $\wedge \circ ({\rm Id}\ot S) = \eta\circ \epsilon$
Since $S^2$ = Id, the Clifford-Hopf algebra is $\ZZ_2$-graded co-commutative. For details, see, e.g., \cite{9611,9908,5aa,6aa}.

Now we turn to the construction of quantum Clifford algebras, comprehensively discussed in, e.g., \cite{5aa, 6aa}. It is well known that standard Clifford algebras are constructed on a quadratic space ($V,g$), where $g$ is a symmetric, bilinear map $g: V^*\times V^* \rightarrow \KK$.
Let $B: V^*\times V^* \rightarrow \KK$ be an arbitrary bilinear
not necessarily symmetric form, in such a way that $(V,B)$ is a reflexive space \cite{6aa}.
In order to define the associated quantum
Clifford algebra, given $u,v,w\in V$, take the exterior algebra $\Lambda(V)$, and using the isomorphism $\tau: V\rightarrow V^\ast$ that induces $B$ by the expression $\tau(u)(v):=B(u,v)$,
the (left) contraction is given by $u\underset{B}{\lrcorner} v = B(u, v)$, satisfying
$w \underset{B}{\lrcorner}(u \wedge v) = (w \underset{B}{\lrcorner} u) \w v + \hat{v} \w  (w
\underset{B}{\lrcorner} u)$, and
$ (u \wedge v) \underset{B}{\lrcorner} w = u \underset{B}{\lrcorner} (v
\underset{B}{\lrcorner}
w)$.
Quantum Clifford algebras are denoted as $\cl(V,B)$ and have been investigated
in \cite{5aa,6aa}.
Let now $B = g+A$, where $A = \frac{1}{2}(B-B^T)$.
Denote $u\underset{A}{\lrcorner} v = A(u, v)$ and $u\underset{g}{\lrcorner} v = g(u, v)$. The $B$-dependent
Clifford product $u\underset{B} v$
 can be split
as  $u\underset{B} v = u\underset{g}{\lrcorner} v + u\dot\w v$ \emph{or}
$u\underset{B} v = u\underset{B}{\lrcorner} v + u \w v$,
where $u\dot\w v = u\w v + A(u,v)  = u\w v + u\underset{A}{\lrcorner} v$. Indeed, regarding the product between $u\in V$ and an element $\psi\in \cl(V,B)$  it follows that
\beq\label{ar}
u\underset{B} \psi = u\underset{B}{\lrcorner}\psi + u\wedge \psi = u\underset{g}{\lrcorner} \psi
+ u\underset{A}{\lrcorner} \psi + u\w \psi = u\underset{g}{\lrcorner} \psi
 + u\dot\w \psi
\eeq\noi

The Wick theorem asserts that $\cl(V,B) \simeq  \cl(V,g)$
as $\ZZ_2$-graded Clifford algebras \cite{5aa,6aa}, although the algebras
$\cl(V,B)$ and $\cl(V,g)$ are not isomorphic with respect to the $\ZZ_n$-grading induced from the exterior algebra  underlying structure, which is not preserved when the process of deformation $\cl(V,g)\Ra\cl(V,B)$ is taken into account. It is possible to express every antisymmetric bilinear form as $A(u,v) := F \underset{g}{\lrcorner} (u\wedge v)$ \cite{39}, where
$F\in\Lambda^2(V)$ is appropriately chosen. Defining the
outer exponential of  $F\in\Lambda^2(V)$ as $e_{\w}^F := 1 + F + \frac{1}{2} F\w F + \cdots +
\frac{1}{n!} \w^n F + \cdots$ ($\w^0 F = 1$) which is a finite series when dim $V$ is finite, the Wick isomorphism $\phi$ is given by
\beq
\cl(V,B) = \phi^{-1}(\cl(V,g)) = e_{\w}^{-F} \w \cl(V,g) \w e_{\w}^F = (\cl(V,g), \langle \;\; \rangle_A)
\eeq where $(\cl(V,g), \langle \;\; \rangle_A)$ indicates that the
$\ZZ_n$ grading arising from the exterior algebra underlying structure is now associated
with the $\dot\w$ exterior product in Eq.(\ref{ar}), and not with the original $\w$ exterior product.
The relation between the $\w$ and the $\dot\w$-grading is given by $u\dot\w v = A(u, v) + u\w v$, showing that the $\ZZ_n$-grading is not preserved
via the Wick isomorphism.

When $V=\RR^{p,q}$, consider the Atiyah-Bott-Shapiro mod 8
index theorem, in particular the case $\cl_{p,q} \simeq
\cl_{p-1,q-1} \otimes \cl_{1,1}.$ We briefly recall this construction obtained, e.g., in \cite{5aa,6aa}.
 According to the $V$ can be split orthogonally with respect to $g$, as
$V = \RR^{p,q} = N_{p-1,q-1} \perp_g M_{1,1}.$
If one applies the Clifford map $\gamma : \RR^{p,q} \Ra \cl_{p,q}$ and
defines its natural restrictions $\gamma^\prime : N_{p-1,q-1} \Ra
\cl_{p-1,q-1},$ $\gamma^{\prime\prime} : M_{1,1} \Ra \cl_{1,1},$  the {periodicity theorem} $
\cl_{p,q} \simeq \cl_{p-1,q-1} \otimes \cl_{1,1}$ is obtained
\cite{BudinichTrautman,Lam73,Maks}.
Using the obvious notation $\cl(\RR^{p,q}) = \cl_{p,q}(g)$ and
introducing the restrictions of the Wick isomorphism $\phi^{-1}\vert_N$ and
$\phi^{-1}\vert_M,$ (here $N=N_{p-1,q-1}$ and $M=M_{1,1}),$ the splitting of $\cl_{p,q}(B)$ is given by the following Theorem \cite{6aa}:
\begin{eqnarray}
\cl_{p,q}(B) &=& \phi^{-1}(\cl_{p,q}(g)) =
\phi^{-1}\left[
\cl_{p-1,q-1}(g\vert_N)\otimes\cl_{1,1}(g\vert_M)
\right]=\cl_{p-1,q-1}(B\vert_N)(\phi^{-1}\otimes)\cl_{1,1}(B\vert_M) \nonumber\\
&=&\cl_{p-1,q-1}(B\vert_N)\,\otimes_{\phi^{-1}}\, \cl_{1,1}(B\vert_M).\label{tc}
\end{eqnarray} The last tensor product $\otimes_{\phi^{-1}}$ is not braided \cite{5aa, 6aa}.
This expression will be used to obtain the $\kappa$-Poincar\'e algebra as a deformation of a subgroup
of the conformal transformations. Note also that the 6-tuple $(\Lambda(V), B, \gamma, \epsilon, \Delta, S)$
is called the quantum Clifford-Hopf algebra.

\section{Conformal maps in the Clifford algebra arena}
In this Section we briefly review some results presented in \cite{e1,ma1}.
Let $\{\g_0, \g_1, \g_2, \g_3\}$ denote an orthonormal frame field in a Lorentzian 4-dimensional spacetime,  satisfying $
\g_\mu\cdot\g_\nu = \me (\g_\mu\g_\nu + \g_\nu\g_\mu) = g_{\mu\nu}$.
The pseudoscalar $\g_5:=\g_{0123}$ satisfies $(\g_5)^2 = -1$ and $\g_\mu\g_5 = -\g_5\g_\mu$.
          					
The algebras $\cl_{2,4}$ and $\cl_{4,1}$ are crucial to consider conformal transformations, i.e., it can be shown that inversions, involutions, dilatations, translations, rotations, transvections and diversions  \cite{hest} can be expressed in terms of the Clifford algebra over the space $\RR^{4,1}$ \cite{e1,e2}.

 The groups
 ${\rm Spin}_+(2,4) = \{R \in \cl^+_{2,4} \;|\;R\tilde{R} = 1\}$ and $
\${\rm pin}_+(2,4) = \{D \in \cl_{4,1} \;|\;D\bar{D} = 1\}
$ can be also
 defined. The inclusion  $
\${\rm pin}_+(2,4)\hko\cl_{2,4}^+ \simeq\cl_{4,1}\simeq\CC\otimes\cl_{1,3}$ follows from the definition. In particular, the Spin groups are useful in the definition of conformal transformations.
 Given the quadratic space $\RR^{p,q}$, consider the injective map
$\varkappa: \RR^{p,q}\rightarrow\RR^{p+1,q+1}$ given by $x\mapsto \varkappa(x) = (x, x\cdot x, 1) = (x, \lambda, \mu)$.
The image of $\RR^{p,q}$ is a subset of the quadric $Q\hko\RR^{p+1,q+1}$, described by the equation
$x\cdot x - \lambda\mu = 0$, the so-called Klein absolute.
 The map $\varkappa$ induces an injective map from  $Q$ in the projective space $\RR\PP^{p+1,q+1}$.
Besides, $Q$ is compact and defined as the conformal compactification ${\widehat{\RR^{p,q}}}$ of $\RR^{p,q}$,  homeomorphic to $(S^p\times S^q)/\ZZ_2$ \cite{ort,ma4,ma5}.
In the very particular case where $p = 0$ and $q = n$, the quadric is homeomorphic to the $n$-sphere $S^n$,
the compactification of  $\RR^n$ via the addition of a point at infinity.
There also exists an injective map $s:\RR\oplus\RR^3 \ri \RR\oplus\RR^{4,1}$ defined by $v\mapsto s(v)$ = {\footnotesize$\bx v&v\bar{v}\\1&{\bar{v}}\ex$}.
The following theorem is introduced by Porteous \cite{port,ort}:
\medbreak
{\bf{Theorem}} {\it $\vvn$ {\rm (i)} the map $\varkappa: \RR^{p,q}\ri \RR^{p+1,q+1};\; x\mapsto (x,  x\cdot x, 1)$, is an isometry.$\;\;{\rm (ii)}$ the map $\pi:Q \ri \RR^{p,q};\;(x, \lambda, \mu)\mapsto x/\mu$ defined where $\lambda\neq 0$ is conformal.
$\;\;{\rm (iii)}$ if $U:\RR^{p+1,q+1}\ri \RR^{p+1,q+1}$ is an orthogonal map, the map
 $\Omega = \pi\circ U \circ \varkappa:\RR^{p,q}\ri \RR^{p,q}$ is conformal.} $\vbn$
\medbreak
\noi The application  $\Omega$ maps  conformal spheres onto conformal spheres\footnote{which can be {quasi-spheres} or hyperplanes.}.
A quasi-sphere is a submanifold of $\RR^{p,q}$, defined by the equation
$
a\;x\cdot x + b\cdot x + c = 0,\;\; a, c\in \RR,\;\;b\in\RR^{p,q},
$ From the assertion $(iii)$ of the theorem above, we see that $\pm U$ induce the same conformal transformation
in $\RR^{p,q}$. The  { conformal group} is defined as
${\rm Conf}(p,q) \simeq  {\rm O}(p+1,q+1)/\ZZ_2$.
The component of Conf$(1,3)$ connected to the identity Conf$_+(1,3)$
denotes the M\"obius group of $\RR^{1,3}$. Besides, SConf$_+$(1,3) denotes the component connected to the identity, time-preserving and future-pointing.

Now, consider the basis $\{\varepsilon_{\BA}\}_{\BA = 0}^5$ of $\RR^{2,4}$ that satisfies the relations
$\vcx_0^2 = \vcx_5^2 = 1$, $\vcx_1^2 = \vcx_2^2 = \vcx_3^2 = \vcx_4^2 = -1$, $\vcx_{\BA} \cdot \vcx_{\BB} = 0,\;(\BA\neq\BB)$.
Consider also the quadratic space $\RR^{4,1}$, with basis $\{E_A\}_{A = 0}^4$, where
\bege\label{r41} E_0^2 = -1, \quad\quad E_1^2 = E_2^2 = E_3^2 = E_4^2 = 1, \quad\quad E_A\cdot E_B = 0 \quad (A\neq B).
\enge
\noi The $\{E_A\}$ can be obtained from $\{\vcx_{\BA}\}$ by the isomorphism
\beq\label{pli}
\xi: \cl_{4,1} &\rightarrow& \la_2(\RR^{2,4})\nonumber\\
E_A &\mapsto& \xi(E_A) = \vcx_{A}\vcx_5.
\eeq

Given a vector $\alpha = \alpha^{\BA}\vcx_{\BA} \in \RR^{2,4}$, we obtain a paravector  $\mmb = \alpha\vcx_5 = \alpha^AE_A + \alpha^5\in\RR\oplus\RR^{4,1}.$
From the periodicity theorem for this case $\cl_{4,1}\simeq\cl_{1,1}\otimes\clt$
and it is possible to express an element of $\cl_{4,1}$ as a  $2\times 2$ matrix with entries in  $\cl_{3,0}$ \cite{e1,ma6}.

A homomorphism  $\vartheta: \cl_{4,1}\rightarrow\clt$ is defined as
$  E_i \mapsto \vartheta(E_i) = E_iE_0E_4 \equiv \ee_i$. It can be seen that  $\ee_i^2 = 1$, $E_i = \ee_iE_4E_0$ and
                    $E_4 = E_+ + E_-,$ $E_0 = E_+ - E_-,$
where $E_\pm := \me(E_4 \pm E_0)$. Then,
\bege\label{mou}
\mmb = \alpha^5 + (\alpha^0 + \alpha^4)E_+ + (\alpha^4 - \alpha^0) E_- + \alpha^i\ee_i E_4E_0.
\enge
\noi  If we choose $E_4$ and $E_0$ to be represented by
$
E_4 =  {\footnotesize\begin{pmatrix}
              0&1\\1&0\end{pmatrix}}$, $E_0 =  {\footnotesize\begin{pmatrix}
              0&-1\\1&0\end{pmatrix}}$
consequently we have $
E_+ =  {\footnotesize\begin{pmatrix}
              0&0\\1&0\end{pmatrix}},\; E_- =  {\footnotesize\begin{pmatrix}
              0&1\\0&0\end{pmatrix}}, \; E_4E_0 =  {\footnotesize\begin{pmatrix}
              1&0\\0&-1\end{pmatrix}}$ and then the paravector $\mmb\in\RR\op\RR^{4,1}\hko\cl_{4,1}$ in Eq.(\ref{mou}) is represented by
$\mmb =  {\footnotesize\begin{pmatrix}
              \alpha^5 + \alpha^i\ee_i&\alpha^4 - \alpha^0\\\alpha^0 + \alpha^4&\alpha^5 - \alpha^i\ee_i\end{pmatrix}}$.

            The vector $\alpha\in\RR^{2,4}$ is in the Klein absolute, i.e., $\alpha^2 = 0 \Leftrightarrow \mmb{\bar{\mmb}} = 0,$
  since
$\alpha^2 = \alpha\alpha = \alpha 1\alpha
= \alpha\vcx_5^2\alpha
 = \alpha\vcx_5\vcx_5\alpha
 = \mmb{\bar{\mmb}}.
$
We denote
$\lambda = \alpha^4 - \alpha^0,$
 $\mu = \alpha^4 + \alpha^0$.  Using the matrix representation of $\mmb{\bar{\mmb}}$, the entry
 ($\mmb{\bar{\mmb}})_{11}$ of the matrix is given by
 \bege \label{xx}
 (\mmb{\bar{\mmb}})_{11} = x{\bar{x}} -\lambda\mu = 0, \quad\text{where}\;\;\;\;
x := (\alpha^5 + \alpha^i\ee_i)\in \RR\oplus\RR^3\hko\clt.
\enge\noi
If we fix $\mu = 1$, consequently $\lambda = x{\bar{x}}\in \RR$.
This choice does correspond to a projective description, and $\mmb\in\RR\oplus\RR^{4,1}$ can be represented as
 \bege\label{parax}
\mmb =  \left(\bea{cc}x &\lambda\\
 \mu&{\bar{x}}
 \ear\right) = \left(\bea{cc}x &x{\bar x}\\
 1&{\bar{x}}
 \ear\right).\nonumber\enge
  From Eq.(\ref{xx}) it follows that
  $
  (\alpha^5 + \alpha^i\ee_i)(\alpha^5 - \alpha^i\ee_i) = (\alpha^4 - \alpha^0)(\alpha^4 + \alpha^0)$, or similarly
  $
  (\alpha^5)^2 + (\alpha^0)^2 - (\alpha^1)^2 - (\alpha^2)^2 - (\alpha^3)^2 - (\alpha^4)^2 = 0,$
 which is the Klein absolute.

Now, the M\"obius transformations in Minkowski spacetime can be derived.
The matrix $ g = {\footnotesize{\bx a&c\\b&d\ex}}$ is in the group \$pin$_+(2,4)$  if, and only if, its entries $a, b, c, d\in\cl_{3,0}$
satisfy the conditions \cite{Maks,e1}
\beq
(i)&&a{\bar{a}},\; b{\bar{b}}, \;c{\bar{c}},\; d{\bar{d}}\in\RR,\quad\quad (ii)\;\;a{\bar{b}},\; c{\bar{d}} \;\in\RR\oplus\RR^3,\quad\quad
(iii)\;\;av{\bar{c}} + c{\bar{v}}{\bar{a}},\;\; cv{\bar{d}} + d{\bar{v}}{\bar{c}}\in \RR, \quad\forall v\in\RR\oplus\RR^{3},\nonumber\\
(iv)&&av{\bar{d}} + c{\bar{v}}{\bar{b}}\in \RR\oplus\RR^{3}, \quad\forall v\in \RR\oplus\RR^{3},\quad\quad
(v)\;\;a{\tilde{c}} =  c{\tilde{a}},\; b{\tilde{d}} = d{\tilde{b}},\qquad\quad(vi)\;\; a{\tilde{d}} - c{\tilde{b}} =   1.
\eeq
\noi Conditions $(i), (ii), (iii), (iv)$ are equivalent to the condition ${\hat{\si}}(g)(\mmb) := g\mmb {\tilde{g}}\in\RR\oplus\RR^{4,1}, \;\forall \mmb\in \RR\op\RR^{4,1}$,
where ${\hat{\si}}: \${\rm pin}_+(2,4)\ri {\rm SO}_+(2,4)$ is the twisted adjoint representation. For details, see, e.g. \cite{e1,e2}.
 Conditions $(v), (vi)$ express $g{\bar{g}} = 1$, for all $g\in\${\rm pin}_+(2,4)$.

  We have just seen   that a paravector $\mmb\in\RR\op\RR^{4,1}$ is represented as $\mmb =
{\footnotesize{\begin{pmatrix} x&x{\bar{x}}\\1&{\bar{x}}
 \end{pmatrix} =  \begin{pmatrix} x&\lambda\\
 \mu&{\bar{x}}
\ex}}$ where $x\in\RR\op\RR^3$ is a paravector of $\clt$.
  Consider $g\in  \${\rm pin}_+(2,4)
$, or similarly $g\in\cl_{4,1}\simeq\cl_{1,1}\otimes\clt$, i.e., $
  g = {\footnotesize{\begin{pmatrix}  a&c\\
 b&d
\end{pmatrix}}} , \,a,b,c,d\in \clt.
$
 The rotation of  $\mmb\in\RR\op\RR^{4,1}$ is performed by the use of the twisted adjoint representation
 ${\hat{\si}}:\${\rm pin}_+(2,4) \ri {\rm SO}_+(2,4)$, defined as
${\hat{\si}}(g)(\mmb) = g\mmb{\hat{g}}^{-1}= g\mmb{\tilde{g}}$, i.e., the adjoint representation of \$pin$_+$(2,4):
 \bege
 \bx a&c\\
 b&d
 \ex \left(\bea{cc} x&\lambda\\
 \mu&{\bar{x}}
 \ear\right){\widetilde{\left(\bea{cc} a&c\\
 b&d
 \ear\right)}} = \left(\bea{cc} a&c\\
 b&d
 \ear\right) \left(\bea{cc} x&\lambda\\
 \mu&{\bar{x}}
 \ear\right)\left(\bea{cc} {\bar d}&{\bar c}\\
 {\bar b}&{\bar a}
 \ear\right)\nonumber
            \enge Fixing $\mu = 1$, the  paravector $\mmb$ is mapped on
\bege\label{acon}
\left(\bea{cc} a&c\\
 b&d
 \ear\right) \left(\bea{cc} x&x{\bar{x}}\\
 1&{\bar{x}}
 \ear\right)\left(\bea{cc} {\bar d}&{\bar c}\\
 {\bar b}&{\bar a}
 \ear\right) = \Delta\left(\bea{cc} x'& x'{\bar{x'}}\\
 1&{\bar{x'}}
 \ear\right),\;\;\; \text{where}\;\; x' := (ax + c)(bx + d)^{-1},\;\;\Delta := (bx + d)({\overline{bx + d}})\in\RR,\enge
\noi which is conformal \cite{vah,hest}.
From the isomorphisms
$
\cl_{4,1}\simeq \CC\otimes\cle \simeq{\mt M}(4,\CC)$,
elements of \$pin$_+$(2,4) are in the Dirac algebra $\CC\otimes\cle$.
 The conformal maps are expressed by the adjoint representation of \$pin$_+$(2,4), by the matrices presented at Table I \cite{port,Maks,vah,hest}.
\begin{table}[tbp]
\centering
{\footnotesize \begin{tabular}{||r||r||r||}\hline\hline
Conformal Map&Explicit Map&Matrix of $\$$pin$_+(2,4)$\\\hline\hline
Translation&$x\mapsto x + h,\;\; h\in \RR\op\RR^3$ &{\footnotesize{$\left(\bea{cc} 1&h\\
 0&1
 \ear\right)$}}\\\hline
Dilation&$x\mapsto \rho x, \;\rho\in\RR $& {\footnotesize{$
\left(\bea{cc} \sqrt{\rho}&0\\
 0&1/\sqrt{\rho}
 \ear\right)$}}\\\hline
Rotation&$x\mapsto \mmg x {\hat{\mmg}}^{-1},\; \mmg\in \${\rm pin}_+(1,3)$ &{\footnotesize{
$\left(\bea{cc} \mmg&0\\
0&{\hat{\mmg}}
 \ear\right)$}}\\\hline
Inversion&$x\mapsto -{\overline{x}}$&
{\footnotesize{
$\left(\bea{cc} 0&-1\\
1&0
 \ear\right)$}}\\\hline
Transvection& $x\mapsto x + x(hx + 1)^{-1},\;\;h\in\ty$&  {\footnotesize{
$\left(\bea{cc} 1&0\\
h&1
 \ear\right) $}} \\\hline\hline
\end{tabular}\vspace{0.5mm}
\caption{\small Conformal maps as Spin group adjoint representations, via the periodicity theorem.}
\label{table1}}
\end{table}
 This index-free geometric formulation allows to trivially generalize the conformal maps of $\RR^{1,3}$ to the ones of $\RR^{p,q}$,
 if the periodicity theorem of Clifford algebras is used.

Elements of \$pin$_+$(2,4) induce the orthochronous M\"obius transformations. The isomorphisms
$
{\rm Conf}(1,3) \simeq {\rm O}(2,4)/\ZZ_2 \simeq {\rm Pin}(2,4)/\{\pm 1, \pm i\}
$ are constructed in \cite{port} and consequently their respective special subgroups are related by
$
{\rm SConf}_+(1,3) \simeq {\rm SO}_+(2,4)/\ZZ_2 \simeq {\rm \$pin}_+(2,4)/\{\pm 1, \pm i\}.
$ The sequence
$
\${\rm pin}_+(2,4)\simeq {\rm SU}(2,2) \stackrel{2-1}{\longrightarrow} {\rm SO}_+(2,4) \stackrel{2-1}{\longrightarrow} {\rm SConf}_+(1,3)$ is explicitly constructed in \cite{Kl74}.

Now, using the well known result that the Lie algebra $ {\mathfrak{spin}}_+(p,q)$ is the algebra  2-covectors of $\RR^{p,q}$, i.e.,
${(\la_2(\RR^{p,q}),\;[\;,\;]) = {\mathfrak{spin}}_+(p,q)},
$ the Lie algebra of the conformal group and the alternative $\kappa$-deformed Poincar\' e algebra can be now. In what follows the Lie algebra
associated with the conformal group is briefly reviewed.
 The basis $\{E_A\}$ defined by Eq.(\ref{pli}) obviously satisfies Eqs.(\ref{r41}).
An isomorphism between $\cl_{4,1}$ and $\CC \ot \cl_{1,3}$ is defined, denoting  $i = \g_{0123} = E_{01234}$,
by \beq E_0 &\mapsto& -i\g_0, \qquad
E_1 \mapsto -i\g_1,\qquad
E_2 \mapsto -i\g_2,\qquad
E_3 \mapsto -i\g_3,\qquad
E_4 \mapsto -i\g_{0123}.\label{poi}\eeq \noindent

The Lie algebra of \$pin$_+$(2,4) is generated by $\Lambda^2(\RR^{2,4})$ and its relation to the group Conf(1,3) is investigated now.
The generators of Conf(1,3) are defined by:
\bege
P_\mu = \frac{i}{2}(\vcx_\mu\vcx_5 + \vcx_\mu\vcx_4),\qquad
K_\mu = -\frac{i}{2}(\vcx_\mu\vcx_5 - \vcx_\mu\vcx_4),\qquad
D = - \me \vcx_4\vcx_5,\qquad
M_{\mu\nu} =  \frac{i}{2} \vcx_\nu\vcx_\mu.\nonumber
\enge
\noi Using (\ref{poi}), the generators of Conf(1,3) are expressed in terms of the  $\{\g_\mu\}\in\cle$ as \cite{e1}:
\bege
P_\mu = \me(\g_{\mu} + i\g_\mu\g_5),\qquad
K_\mu = -\me (\g_{\mu} - i\g_\mu\g_5),\qquad
D = \me i \g_5,\qquad
M_{\mu\nu} = \me(\g_\nu\wedge\g_\mu).\label{confa}
\enge \noi They satisfy the following relations:
\beq\label{con}
[P_\mu, P_\nu] &=& 0,\quad\quad [K_\mu, K_\nu] = 0,\quad\quad [M_{\mu\nu}, D] = 0,\nonumber\\
 \left[M_{\mu\nu}, P_{\lambda}\right] &=& -(g_{\mu\lambda}P_\nu - g_{\nu\lambda}P_\mu),\qquad\qquad
 \left[M_{\mu\nu}, K_{\lambda}\right] = -(g_{\mu\lambda}K_\nu - g_{\nu\lambda}K_\mu),\nonumber\\
 \left[M_{\mu\nu}, M_{\sigma\rho}\right] &=& g_{\mu\rho}M_{\nu\sigma} +
g_{\nu\si}M_{\mu\rho} - g_{\mu\si}M_{\nu\rho} - g_{\nu\rho}M_{\mu\si},\label{M}\nonumber\\
 \left[P_\mu, K_\nu\right] &=& 2(g_{\mu\nu} D - M_{\mu\nu}),\qquad \left[P_\mu, D\right] = P_\mu,\qquad
 \left[K_\mu, D\right] = -K_\mu.
\eeq \noi The commutation relations above
are invariant under substitution $P_\mu \mapsto - K_\mu$, $K_\mu \mapsto - P_\mu$ and $D\mapsto -D$.
All the above relations are derived essentially from the periodicity theorem that implements the isomorphism
$\cl_{2,4}\simeq \cl_{1,1}\ot \cl_{1,3}$  and also the isomorphism $\cl_{2,4}^+\simeq \cl_{4,1}$. In the next Section
the deformation of the periodicity theorem will be used to derive the $\kappa$-Poincar\'e algebra.

\section{The $\kappa$-Poincar\'e algebra as a quantum Clifford-Hopf algebra}
Usually the standard real $\kappa$-Poincar\'e algebra can be obtained via the commutators of $\mathcal{U}_q$($\mathfrak{so}$(3,2))
\cite{lu2,lu3,8gi,9gi}
and subsequently performing the quantum de-Sitter contraction. Here the aim is to construct an algebra that is equivalent
to the deformed anti-de Sitter algebra $\mathcal{U}_q$($\mathfrak{so}$(3,2)).

Starting from Eqs.(\ref{confa}), they express Eqs.(\ref{con}) in terms of elements in $\cle$. Take the group $\$$pin$_+$(2,4) = $\$$pin$_+({2,4})$($g$), and use the Wick isomorphism $\phi$($\$$pin$_+$(2,4)) = $\$$pin$_+$({2,4})($B$). There is a $F\in\Lambda^2(\RR^{1,3})$ such that the relations (\ref{con}) can be deformed, and using Eqs.(\ref{confa}) and the expression $e^{-F}_\w \psi e^{F}_\w$ that defines the Wick isomorphism
on each $\psi\in\cle$, the $\kappa$-Poincar\'e algebra can be obtained by a suitable particular
construction of the non braided tensor product, in the periodicity theorem of quantum Clifford algebras.
Regarding the deformed tensor product that is not braided, we also obtain a Hopf-algebraic structure
arising from quantum Clifford algebras. The formalism accomplished by Eqs.(\ref{con}) can be turned in
the expressions defining the $\kappa$-Poincar\'e algebra, if the not braided tensor product in Eq.(\ref{tc}) is appropriately chosen.

Denoting $K_\pm = K_1 \pm iK_2:= M_{10}\pm iM_{20}$, $K_3 = M_{30}$, $M_\pm
= M_{23} \pm  iM_{31}$, and $P_\pm = P_2 + iP_1$, and using Eqs.(\ref{confa})
 the $\kappa$-Poincar\'e algebraic sector is presented by the following commutation relations:
\beq
[P_\mu,P_\nu]&=& 0 = [M_{ij},P_0], \qquad [\epsilon_{ijk}M_{ij},P_\ell]=i\,\epsilon_{i\ell r}P_r\nonumber\\
\,[K_3,P_0]&=& \frac{i}{2}\g_3(1+i\g_5),\qquad [K_3,P_2]= \frac{i}{2\kappa}\g_2\g_3(1+i\g_5),\qquad [P_3,K_3]=\frac{i}{2\kappa}
(1+i\g_5)-i\kappa \sinh \left(\frac{\g_0 + i\g_0\g_5}{\kappa}\right)\nonumber\\
\,[K_3,P_1]&=& \frac{i}{2\kappa}\g_3\g_1(1+i\g_5),\qquad [K_\pm,P_0]=\me(\mp \g_2+i\g_1)(1+i\g_5),\nonumber\\
\,[K_\pm,P_2]&=& \mp i\kappa \sinh \left(\frac{\g_0 + i\g_0\g_5}{\kappa}\right) \pm \frac{1}{2\kappa} \g_3(1+i\g_5)\nonumber\\
\,[K_\pm,P_1]&=&  i\kappa \sinh \left(\frac{\g_0 + i\g_0\g_5}{\kappa}\right) - \frac{i}{2\kappa} \g_3(1+i\g_5),\qquad
[K_\pm,P_3]=\mp\frac{1}{2\kappa} \g_3(\g_2\mp i\g_1)(1\pm i\g_5)\nonumber\\
\,[M_+,M_-]&=& \me\g_1\wedge\g_2,\qquad\qquad [M_{12},M_\pm]=\pm \me\g_3(\g_1\pm i\g_2)\nonumber\\
\, [K_+,K_-]&=& -\g_1\wedge\g_2\cosh \left(\frac{\g_0 + i\g_0\g_5}{\kappa}\right) - \sinh \left(\frac{\g_0 + i\g_0\g_5}{\kappa}\right)\nonumber\\
\,[K_\pm,K_3]&=& \pm 1 \pm \frac{\g_0}{4\kappa}(1+i\g_5)\g_3(\g_2-i\g_1) + \frac{1}{8\kappa}\left((i+1)\g_3\wedge \g_0(\g_2+i\g_1)(1+i\g_5)
\right)\nonumber\\
\,[M_\pm,K_\pm]&=& \mp \frac{1}{8\kappa}\g_3(1+i\g_1\g_2)(1\mp 1) (1+i\g_5),\qquad\qquad [M_{12},K_3]=0,\qquad\qquad [M_{12},K_\pm]=\mp\me
(\g_1\pm i\g_2)\wedge\g_0\nonumber\eeq\beq
\,[M_\pm,K_\mp]&=&\left(\mp\g_3+\frac{i}{8\kappa}(1\mp 1)(1-\g_1\g_2)\wedge \g_3\right)(1-i\g_5)+\frac{1}{4}(\mp\g_1\wedge\g_2\pm 2)
\g_3(1+i\g_5)\nonumber\\
\,[M_\pm,K_3]&=&\mp\me(\g_1\pm i\g_2)\wedge\g_0 \pm\frac{1}{8\kappa}(\g_1\wedge\g_2)(\g_1+i\g_2)(1+i\g_5) + \frac{i}{4\kappa}(\g_2\mp i\g_1)
(1+i\g_5).\nonumber
\eeq
The coalgebra sector associated with the above Lie algebra is given by the antipodes
\beq
S(M_{ij})&=&-M_{ij},\qquad S(P_\mu)=-P_\mu,\qquad
S(K_3) = -\me \g_3(1-i\g_5) + \frac{i}{2\kappa}\g_3(1+i\g_5)+ \frac{\g_1}{2\kappa},\nonumber\\
S(K_\pm)&=& -\me(\g_1+i\g_2)\wedge\g_0 \pm \frac{1}{2\kappa}(\g_2\mp i\g_1) \mp \frac{i}{4\kappa}(\g_1\mp i\g_2)(1+i\g_5)\nonumber
\eeq
and by the coproducts
\beq \Delta(M_{ij}) &=& \me (\g_i\wedge\g_j\otimes 1 + 1\otimes \g_i\wedge\g_j)\nonumber\\
\Delta(K_3)&=& -\frac{\g_3}{2}(1-i\g_5)\ot \left(1+\frac{\g_0}{2\kappa}(1+i\g_5)\right) + \left(1-\frac{\g_0}{2\kappa}(1+i\g_5)\right)
\ot \frac{\g_3}{2}(1-i\g_5)\nonumber\\
&& + \frac{1}{4\kappa}\left(1-\frac{\g_0}{2\kappa}(1+i\g_5)\right)(\g_2\wedge\g_3 \ot \g_2(1+i\g_5))\nonumber\\
\Delta(K_\pm)&=&-\me(\g_1\pm i\g_2)\wedge\g_0 \ot \left(1+\frac{\g_0}{2\kappa}(1+i\g_5)\right) +
\left(1-\frac{\g_0}{2\kappa}(1+i\g_5)\right)\ot  \frac{\g_3}{2}(1-i\g_5)\nonumber\\
&& + \frac{1}{2\kappa} \left(1-\frac{\g_0}{2\kappa}(1+i\g_5)\right) (\g_2\mp i\g_1)\ot \me (\g_1\wedge\g_2) \left(1+\frac{\g_0}{2\kappa}(1+i\g_5)\right)- \left(1-\frac{\g_0}{2\kappa}(1+i\g_5)\right) \me(\g_2\mp i\g_1)\ot (\g_2\mp i\g_1)\nonumber\\
&&\mp\frac{i}{2\kappa} \left(1+\frac{\g_0}{2\kappa}(1+i\g_5)\right)\g_3\wedge(\g_1\pm i\g_2)\ot \g_3(1+i\g_5)\nonumber\\
\Delta(P_i)&=& \me\left(\g_i(1+i\g_5) \ot (1 +  \g_0(1+i\g_5)) + (1 +  \g_0(1+i\g_5)) \ot \g_i(1+i\g_5)\right)\nonumber
\eeq
Using Eqs.(\ref{confa}) and defining
\beq
\mathring{K}_3&=& \me\g_{30} -\frac{i}{16\kappa}\g_3(1-4i+\g_{12})(1 + i\g_5),\quad
\mathring{K}_\pm = \me\left(i\g_0 + \frac{1}{2\kappa} (\pm 1-\frac{i}{4})\right)\frac{1}{2}(1\pm i\g_5)(\g_1\pm i\g_2),\nonumber
\eeq \noi these generators satisfy, besides Eq.(\ref{M}), the following commutation relations:
\beq
[M_j,\mathring{K}_k]&=&i\epsilon_{jkl}\mathring{K}_l,\qquad
[\mathring{K}_k,P_0]=\frac{i}{2} \g_k (1+i\g_5),\qquad
[\mathring{K}_j,P_k]=i\,\kappa\, g_{kj}\,\sinh \left(\frac{\g_0+ i\g_0\g_{5}}{\kappa}\right),\nonumber\\ {}
[\mathring{K}_j,\mathring{K}_k]&=&-i\g_j\wedge\g_k\cosh \left(\frac{\g_0 + i\g_0\g_{5}}{\kappa}\right) - \frac{1}{4\kappa^ 2}
 \epsilon_{rpq}\g_k \g^r \g_5 \g^p\wedge\g^q,\nonumber\eeq\noi
 A $\kappa$-deformation of the Poincar\' e algebra is obtained, and it coincides with \cite{6gi}.
It can be forthwith shown that those equations satisfy all the well known $\kappa$-Poincar\' e algebra commutation relations  given in \cite{lu1}.
The respective coalgebra sector is described by the coproducts
\beq
\Delta(M_{ij}) &=& \me (\g_i\wedge\g_j\otimes 1 + 1\otimes \g_i\wedge\g_j),\qquad\qquad \Delta(P_0)= \me\left(\g_0(1+i\g_5) \ot 1 + 1\ot \g_0(1+i\g_5)\right)\nonumber\\
 \Delta(\mathring{K}_i) &=& -\frac{1}{4}\left([1+(\g_i+\g_0)(1-i\g_5)] \ot [1+(\g_i+\g_0)(1-i\g_5)]\right)\nonumber\\
&& + \frac{i}{2\kappa}\epsilon_{ijk}\left[\g_j(1+i\g_5)\ot \g_i \g_j (1 +  2\kappa^{-1}\g_0(1+ i\g_5)) + \g_i \g_j
(1 -  2\kappa^{-1}\g_0(1+ i\g_5))\ot \g_j(1+i\g_5)\right]\nonumber\\
\Delta(P_i)&=& \me\left(\g_i(1+i\g_5) \ot (1 +  \g_0(1+i\g_5) + (1 +  \g_0(1+i\g_5) \ot \g_i(1+i\g_5)\right)\nonumber
\eeq\noi The counits are given by $\epsilon(M_{\mu\nu})=\epsilon(P_\mu)=\epsilon(\mathring{K}_\mu)=0$,
and the antipodes are given by
$
S(M_{\mu\nu})=\g_{\mu\nu},\;\, S(P_\mu)=P_\mu, \;\, S(\mathring{K}_j)=\g_j\left(\me(1-i\g_5)+\frac{3i}{4\kappa}\me(1+i\g_5)\right).$
 These results can be compared with \cite{9gi}.

Now the quantum $\kappa$-deformed Poincar\'e symmetries are formulated
in modified bicrossproduct basis with classical Lorentz subalgebra, together with
the respective Hopf algebra relations. A realization of the algebraic sector of a
quantum $\kappa$-deformed Poincar\'e algebra can be obtained via
generators $M_{\mu\nu}, K_i, P_\mu$ that satisfy the following relations (besides Eq.(\ref{M})):
\beq
\left[M_{\mu\nu}, K_{\lambda}\right] &=& -(g_{\mu\lambda}K_\nu - g_{\nu\lambda}K_\mu),\qquad \left[K_j,P_0\right]= - i\,\g_j \left(1-\g_0\frac{1 - i\g_{5}}{2\kappa}\right)\frac{1}{2}(1+i\g_5)\nonumber\\
\left[P_k, K_j\right]&=& i\, g_{kj}\g_0 \left(\frac{1 + i\g_{5}}{2}\right)(1- 4\kappa^{-1}) + 2\kappa^{-1}\g_0
\left(1- \frac{1 + i\g_{5}}{4\kappa}\right) \g_j\g_k \g_5\nonumber\eeq\noi The commutation relations involving the rotations $M_{ij}$  and the time
translation $P_0$ are classical, the momenta commute, and quantum deformation appears
only when the boosts are involved.
The respective coalgebra sector is described by the coproducts
\beq
\Delta(M_{ij}) &=& \me (\g_i\g_j\otimes 1 + 1\otimes \g_i\g_j),\qquad\qquad \Delta(P_0)= \me\left(\g_0(1+i\g_5) \ot 1 + 1\ot \g_0(1+i\g_5)\right)\nonumber\\
 \Delta(K_i) &=& -\me\left(\g_i(1+i\g_5)\otimes 1 + (4\kappa^{-1})(1-\g_0(1+i\g_5))\otimes
\g_i(1+i\g_5)\right) + \frac{1}{2\kappa}\epsilon_{ijk}\,\g_j(1+i\g_5)\ot \g_i \g_j\nonumber\\
\Delta(P_i)&=& \me\left(\g_i(1+i\g_5) \ot (1 +  \g_0(1+i\g_5) + (1 +  \g_0(1+i\g_5) \ot \g_i(1+i\g_5)\right).\nonumber
\eeq
In addition, the Wick isomorfirsm can also be applied to the elements exhibited in Table I, also letting 
the Wick isomorphism deformation accomplishment explicit.

\section{Concluding remarks and outlooks}

In this paper we have been concerned to reveal the algebraic aspects
of quantum Clifford algebras as a natural arena to construct a $\kappa$-deformed Poincar\'e algebra
from the original  Poincar\'e algebra, using the Wick theorem.
After reviewing the algebra of conformal transformations and expressing it in terms of the spacetime algebra and 
the Atiyah-Bott-Shapiro periodicity theorem, some algebraic and coalgebraic aspects associated to conformal maps have been conceived in the context of quantum Clifford algebras.
In this scenario, the Lie algebra associated with the conformal group and an alternative $\kappa$-deformed Poincar\'e algebra accrue solely in terms
of elements of the spacetime algebra  $\cle$, without any reference to representations.
The $\kappa$-deformed Poincar\'e algebra is represented in terms of quantum Clifford algebras, and the coalgebraic sector is obtained.
A possible perspective is to investigate our previous results \cite{alex1,alex2} in this context.

As the quantum $\kappa$-Poincar\'e algebra corresponds to DSR in the same way as the standard Poincar\'e algebra is related
to special relativity, it has been argued that the knowledge of this quantum algebra is fundamental in the DSR theory, since the $\kappa$-Poincar\'e algebra is a quantum nonlinear algebra, and it can generate nonlinear transformations
among momenta. 
Here we have presented another distinct basis in terms of the generators of spacetime algebra $\cle$, for which the construction is motivated by the results obtained in \cite{9gi}. 
Formally, distinct basis of the algebra related by analytical mappings of momenta are completely equivalent, and the existence of physical equivalence remains an open problem.
For more details and examples see, e.g., \cite{smolin}.

\begin{acknowledgments}
A. E. B. would like to thank the financial support from the Brazilian
Agencies FAPESP grant 08/50671-0 and CNPq grant 300627/2007-6. R. da Rocha thanks CNPq 304862/2009-6 for financial support.
\end{acknowledgments}

\begin{thebibliography}{99}



\bibitem{lu2} J. Lukierski, A. Nowicki, H. Ruegg, and V. N. Tolstoy, \emph{Phys. Lett.} {\bf B264} 331 (1991).
\bibitem{majid} S. Majid and H. Ruegg, \emph{Phys. Lett.} {\bf B334} (1994) 348 [{\tt arXiv:hep-th/9405107v2}].
\bibitem{lu3} J. Lukierski, H. Ruegg, and W. J.
Zakrzewski, \emph{Ann. Phys.} {\bf 243} (1995) 90 [{\tt arXiv:hep-th/9312153v1}].
\bibitem{lu1} J. Lukierski, A. Nowicki, and H. Ruegg,
\emph{Phys. Lett.} {\bf B293} (1992) 344.
\bibitem{6gi} S. Giller, P. Kosinski, M. Majewski, P. Maslanka, and J. Kunz, \emph{Phys.
Lett.} {\bf B 286} (1992) 57.
\bibitem{7gi}  S. Zakrzewski, \emph{J. Phys.} {\bf A27} (1994) 2075.
\bibitem{8gi}J.  Lukierski  and H. Ruegg, \emph{Phys. Lett.} {\bf B329} (1994) 189 [{\tt arXiv:hep-th/9310117v1}].
\bibitem{9gi} J. Lukierski, A. Nowicki, and H. Ruegg, \emph{$D=4$ quantum Poincar\'e algebras and finite difference with time derivatives},
 in \emph{Spinors, Twistors, Clifford Algebras
and Quantum Deformations}, Z. Oziewicz, B. Jancewicz, and A. Borowiec (Eds.), Kluwer, Dordrecht 1993.
\bibitem{kappa} E. Celeghini, R. Giachetti, E. Sorace, and M. Tarlini, \emph{J. Math. Phys.} {\bf 31}
(1990) 2548; J. Math. Phys. {\bf 32} (1991) 1155.
\bibitem{ame1} G. Amelino-Camelia, \emph{Int. J. Mod. Phys.} {\bf D 11} (2002) 35 [{\tt arXiv:gr-qc/0210063v1}];
 \emph{Phys. Lett.} {\bf B510} (2001) 255 [{\tt arXiv:hep-th/0012238v1}].
\bibitem{bruno}N. R. Bruno, G. Amelino-Camelia, and J. Kowalski-Glikman, \emph{Phys. Lett.} {\bf B522} (2001) 133 [{\tt arXiv:hep-th/0107039v1}].
\bibitem{glik}J. Kowalski-Glikman, \emph{Phys. Lett.} {\bf A286} (2001) 391 [{\tt arXiv:hep-th/0102098v2}].
\bibitem{now1}
J. Kowalski-Glikman and S. Nowak,  \emph{Int. J. Mod. Phys.} {\bf D12} 299
(2003) [{\tt arXiv:hep-th/0204245v1}]; \emph{Phys. Lett.} {\bf B539} 126
(2002) [{\tt arXiv:hep-th/0203040v1}].
\bibitem{port} I. R. Porteous, {\it Clifford Algebras and the Classical Groups},  (Cambridge Studies in Advanced Mathematics),
Cambridge Univ. Press, Cambridge 1995.
\bibitem{Kl74} F. Klotz,
 \emph{J. Math. Phys.} {\bf 15} (1974) 2242.
 \bibitem{hest} D. Hestenes, \emph{Acta Appl. Math.} {\bf 23} (1991) 65.
\bibitem{e1} R. da Rocha and J. Vaz, Jr., \emph{Int. J. Geom. Meth. Mod. Phys.} {\bf 4} (2007) 547-576
 [{\tt arXiv:math-ph/0412074v2}];  [{\tt arXiv:math-ph/0412075v1, arXiv:math-ph/0412076v1}].
\bibitem{e2} W. A. Rodrigues Jr., R. da Rocha, and J. Vaz, Jr., \emph{Int. J. Geom. Meth. Mod. Phys.} {\bf 2} (2005) 305-357
[{\tt arXiv:math-ph/0501064v6}].
\bibitem{e3} R. da Rocha and J. Vaz, Jr., \emph{Int. J. Theor. Phys.} {\bf 46} (2007) 2464 [{\tt arXiv:0710.0832v1}].
\bibitem{e4} V. V. Fernandez, W. A. Rodrigues Jr., A. M. Moya, and R. da Rocha, \emph{Int. J. Geom. Meth. Mod. Phys.} {\bf 4} (2007) 1159-1172 [{\tt arXiv:math/0502003v5}].
\bibitem{ma1} R. da Rocha and W. A. Rodrigues, Jr., \emph{Annalen der Physik} {\bf 19} (2010) 06-34 [{\tt 	arXiv:0811.1713v7  [math-ph]}], [{\tt 	arXiv:0912.2127v1  [math-ph]}].
\bibitem{ma4} R. da Rocha and W. A. Rodrigues, Jr., \emph{Adv. Appl. Clifford Alg.} {\bf 18} (2008)  351-367 [{\tt arXiv:math-ph/0510026v2}].
\bibitem{ma5} R. 	da Rocha and J. Vaz, Jr., \emph{J. Algebra} {\bf 301} (2006) 459-473  	[{\tt arXiv:math-ph/0603053v1}].
\bibitem{ma6} R. da Rocha and J. Vaz, Jr.,  \emph{Int. J. Geom. Meth. Mod. Phys.} {\bf 3} (2006) 1359-1380 [{\tt arXiv:math-ph/0605009v1}].

\bibitem{alex1} A. E. Bernardini and R. da Rocha, \emph{Phys. Rev.} {\bf D75} (2007) 065014 [{\tt arXiv:hep-th/0701094v2}].
\bibitem{alex2}  A. E. Bernardini, \emph{Phys. Rev.} {\bf D75} (2007) 097901 [{\tt 	arXiv:0706.3932v2  [hep-ph]}].
\bibitem{ma3} A. E. Bernardini and R. da Rocha,  \emph{Europhys. Letters} {\bf 81} (2008)  40010 [{\tt arXiv:hep-th/0701092v2}].

\bibitem{smolin} J. Magueijo and L. Smolin, \emph{Phys. Rev. Lett.} {\bf 88} (2002) 190403 [{\tt arXiv:hep-th/0112090v2}].
\bibitem{zumino} O. Ogievetski, W. B. Schmidtke, J. Wess, and B. Zumino, \emph{Comm. Math.}
Phys. {\bf 150} (1992) 495.
\bibitem{g1} S. Giller, P. Kosinski, M. Majewski, P. Maslanka, and J. Kunz, \emph{Phys.
Lett.} {\bf B286} (1992) 57.
\bibitem{g2} V. K. Dobrev, \emph{J. Phys.} {\bf A26} (1993) 1317.
\bibitem{g3} H. Bacry, \emph{J. Phys.} {\bf A26}  (1993) 5413; Phys. Lett. {\bf 306B} (1993) 41.
\bibitem{g4} S. Majid, \emph{J. Math. Phys.} {\bf 34} (1993) 2045 [{\tt arXiv:hep-th/9210141v1}].
\bibitem{g5}  M. Chaichian and A. Demitchev, \emph{Phys. Lett.} {\bf B304}  (1993) 220.
\bibitem{h1} S. Majid and H. Ruegg, \emph{Phys. Lett.} {\bf B334}  (1994) 348 [{\tt arXiv:hep-th/9405107v2}].
\bibitem{h2} A. Ballesteros, F. J. Herranz, M. A. del Olmo, and M. Santander, \emph{Phys.
Lett.} {\bf B351}  (1995) 137.
\bibitem{h3} J. A. de Azcarraga and J. C. Perez Bueno, \emph{J. Math. Phys.} {\bf 36} (1995) 6879 [{\tt q-alg/9505004v3}].
\bibitem{h31} J. A. de Azcarraga and F. Rodenas,  \emph{J. Phys.} {\bf A29} (1996) 1215 [{\tt arXiv:q-alg/9510011v2}].
\bibitem{h32}  J. A. de Azcarraga, P. P. Kulish, and F. Rodenas, \emph{Lett. Math. Phys.} {\bf 32} (1994) 173
[{\tt arXiv:hep-th/9309036v1]}.
\bibitem{i1} L. Castellani, Phys. Lett. {\bf B327} (1994) 22 [{\tt arXiv:hep-th/9402033v2}].
\bibitem{i3} P. Kosinski and P. Maslanka, in \emph{From Quantum Field Theory to Quantum
Groups}, Ed. B. Jancewicz and J. Sobczyk, World Sc., 1996, p. 41.
\bibitem{61} S. L. Woronowicz and S. Zakrzewski, \emph{Comp. Math.} {\bf 90} (1994) 211.
\bibitem{i2} P. Podles and S. L. Woronowicz, \emph{Comm. Math. Phys.} {\bf 178} (1996) 61.
\bibitem{9611} B. Fauser, \emph{Adv. Appl. Clifford Alg.} {\bf  10} (2000) 173
[{\tt arXiv:hep-th/9611069v1}].
\bibitem{9908} R. Ablamowicz, B. Fauser, in "Clifford Algebras and their Applications in Mathematical Physics", R. Ab{\l}amowicz, B. Fauser eds. Birkh\"auser, Boston 2000, pp. 245-268 [{\tt arXiv:math/9908062v1 [math.QA]}].
\bibitem{a36} A. J. Hahn, \emph{Quadratic Algebras, Clifford Algebras, and Arithmetic Witt Groups}, Springer, New York 1994.
\bibitem{cox} H. S. M. Coxeter and W. O. J. Moser, \emph{Generators and Relations for Discrete
Groups}, Springer Verlag, Berlin 1980.
\bibitem{kos} K. Przanowski [{\tt arXiv:q-alg/9606022v1}].
\bibitem{drin} V. G. Drinfeld, in \emph{Quantum Groups}, Proc. Int. Congress of Mathematics,
Berkeley, 1986, p. 798.
\bibitem{fadeev} L. Faddeev, N. Resetikhin, and L. Takhtajan, \emph{Alg. Anal.} {\bf 1} (1990) 178.
\bibitem{majid1} S. Majid, \emph{Foundations of Quantum Group Theory}, Cambridge University Press, Cambridge, 1995.
\bibitem{q1} M. Dubois-Violette and G. Launer, \emph{Phys. Lett.} {\bf 245B} (1990) 175.
\bibitem{q2} E. Demidov,  Yu. I. Manin, E. E. Mukhin, and D. V. Zhdanovich, \emph{Progr. Theor. Phys.
Suppl.} {\bf 102} (1990) 203.
\bibitem{q3} S. Zakrzewski, \emph{Lett. Math. Phys.} {\bf 22} (1991) 287.
\bibitem{q4} H. Ewen, O. Ogievetsky, and J. Wess, \emph{Lett. Math. Phys.} {\bf 22}  (1991) 297.
\bibitem{q5} C. Ohn, \emph{Lett. Math. Phys.} {\bf 25} (1992) 85.
\bibitem{38} M. E. Sweedler, \emph{Hopf Algebras}, W. A. Benjamin, Inc., New York 1969.
\bibitem{5aa} B. Fauser and R. Ablamowicz: in Clifford Algebras and their Applications in Mathematical
Physics, (Eds. R. Ablamowicz, B. Fauser), Birkhauser, Boston 2000, p. 347 [{\tt 	arXiv:math/9911180v2 [math.QA]}].
\bibitem{6aa} B. Fauser, \emph{J. Phys.} {\bf A}: \emph{Math. Gen.} {\bf 32} (1999) 1919 [{\tt arXiv:hep-th/0007032v1}]; \emph{J. Math. Phys.} {\bf 39} (1998) 4928 [{\tt arXiv:hep-th/9710186v1}]; J. Math. Phys. {\bf 37} (1996) 72-83 [{\tt 	 arXiv:hep-th/9504055v1}]; .
\bibitem{BudinichTrautman} P. Budinich, and A. Trautmann; {\it The Spinorial
Chessboard}, Trieste Notes in Physics, Springer-Verlag, Berlin 1988.
\bibitem{Lam73} T. Y. Lam, {\it The Algebraic Theory of Quadratic Forms}, The Benjamin/Cummings Publishing Company, Reading 1973.
\bibitem{Maks} J. Maks, {\it Modulo (1,1) periodicity of Clifford algebras
and the generalized (anti-) M\"obius transformations\/} Thesis, TU Delft
1989.
\bibitem{39} D. Hestenes and G. Sobczyk, \emph{Clifford Algebra to Geometric Calculus}, Reidel,
Dordrecht 1984.
\bibitem{ort} I. R. Porteous, \emph{A tutorial on conformal groups},  \emph{Generalizations of Complex Analysis and Their Applications in Physics (Warsaw/Rynia, 1994)},  Banach Center Publ. {\bf 37} 137-150, Polish Acad. Sci., Warsaw 1996.
\bibitem{vah}  K. T. Vahlen, \emph{Math. Ann.} {\bf 55}, 585-593 (1902) .



\end{thebibliography}
\end{document}